\documentclass[useAMS,usenatbib]{mn2e}

\usepackage[latin1]{inputenc}
\usepackage[T1]{fontenc}
\usepackage{mathptmx}
\usepackage[scaled=.90]{helvet}
\usepackage{courier}

\usepackage{xcolor}
\usepackage{hyperref}
\hypersetup{colorlinks=true,citecolor=blue, urlcolor=blue, linkcolor=blue}
\usepackage{graphicx}
\usepackage[]{units}
\usepackage{multirow}
\usepackage{xfrac}
\usepackage{aas_macros}
\usepackage{etoolbox}
\usepackage{mnrasArxivFix}

\title[Constraining DM in dSphs with stellar streams]{Constraining the distribution of dark matter in dwarf spheroidal galaxies with stellar tidal streams}

\author[Rapha\"el Errani, Jorge Pe\~narrubia \& Giuseppe Tormen]{Rapha\"el Errani$^{1,2,}$\thanks{E-mail:
raer@roe.ac.uk}, Jorge Pe\~narrubia$^{2}$ \& Giuseppe Tormen$^{1}$\\
{$^1$ Dipartimento di Fisica e Astronomia `Galileo Galilei', Universit\`a degli Studi di Padova, Vicolo Osservatorio 3, 35122 Padova, Italy}\\
{$^2$ Institute for Astronomy, University of Edinburgh, Royal Observatory, Blackford Hill, Edinburgh EH9 3HJ, UK}
}

\begin{document}

\date{Accepted 2015 January 20. Received 2015 January 14; in original form 2014 November 20}

\pagerange{\pageref{firstpage}--\pageref{lastpage}} \pubyear{2015}

\maketitle

\label{firstpage}

\begin{abstract}

We use high-resolution $N$-body simulations to follow the formation and evolution of tidal streams associated to dwarf spheroidal galaxies (dSphs). The dSph models are embedded in dark matter (DM) haloes with either a centrally-divergent `cusp', or an homogeneous-density `core'.
In agreement with previous studies, we find that as tides strip the galaxy the evolution of the half-light radius and the averaged velocity dispersion follows well-defined tracks that are mainly controlled by the amount of mass lost. Crucially, the evolutionary tracks behave differently depending on the shape of the DM profile: at a fixed remnant mass, dSphs embedded in cored haloes have larger sizes and higher velocity dispersions than their cuspy counterparts. 
The divergent evolution is particularly pronounced in galaxies whose stellar component is strongly segregated within their DM halo and becomes more disparate as the remnant mass decreases. 
Our analysis indicates that the DM profile plays an important role in defining the internal dynamics of tidal streams. We find that stellar streams associated to cored DM models have velocity dispersions that lie systematically above their cuspy counterparts. Our results suggest that the dynamics of streams with known dSph progenitors may provide strong constraints on the distribution of DM on the smallest galactic scales.

\end{abstract}



\section{Introduction}
It is widely believed that the microscopic properties of dark matter (DM) particles are reflected upon the distribution of DM on the smallest galactic scales \citep[e.g.][]{TremaineGunn1979, Gilmore2007}. Being the smallest and most dark matter-dominated galaxies in the known Universe, dwarf spheroidal galaxies (dSphs) place strong constraints on the DM particle mass
\citep[e.g.][]{Boyarsky2009,Maccio2012,Lovell2014} as well as on its cross section \citep{Burkert2000,Loeb2011,Vogelsberger2012,Vogelsberger2014,Rocha2013}.
Furthermore, with luminosities in the range of $10^4-10^8L_\odot$ \citep{McConnachie2012}, these galaxies sample the faintest end of the galaxy luminosity function, providing a testbed for theoretical models that aim to understand the dynamical interplay between baryons and the DM haloes \citep[e.g.][]{Navarro96,Pontzen2012,Pontzen2014,Penarrubia2012,Brooks2014,DiCintio2014,Nipoti2014}.

However, measuring the distribution of DM from the internal kinematics of dSphs has proven to be challenging. Their low surface brightness plus their relatively large heliocentric distances limit the number of stellar tracers for which kinematic information can be obtained and also reduce the phase-space information to quantities projected along the line of sight. In addition, as foreground contaminants tend to dominate in number in the outskirts of these galaxies most spectroscopic surveys suffer from strong sampling biases that severely complicate the dynamical analysis of these objects. Given the number and graveness of these shortcomings it is perhaps not surprising that different methods that aim to measure the mass profile of Milky Way dSphs yield contradictory results. For example, whereas some studies find evidence for constant-density DM cores \citep{Battaglia2008,Walker2011,Amorisco2012}, other methods often applied to the same data suggest centrally-divergent `cusps' \citep{Richardson2014}
or remain inconclusive \citep{Breddels2013,Strigari2014}.

Recent numerical studies have shown that the dynamical properties of tidal streams are sensitive to the internal mass distribution of the progenitor system \citep{Kupper2008,Kupper2012,Penarrubia2006,Penarrubia2010,Eyre2011}.
Using high-resolution N-body simulations we show in this paper that the distribution of dark matter in dSphs plays an important role in defining the orbits of tidally stripped stars, and that the dynamics of stellar streams provide constraints on the dark matter halo profile of dSphs which are complementary to those inferred from the internal kinematics of the progenitor galaxy.
The Letter is organised as follows: Section 2 describes the $N$-body dSph and analytical host galaxy models as well as the numerical setup. In Section 3 we analyse how cuspy and cored dSphs evolve under tidal stripping and introduce the evolutionary tracks. In Section 3 we present our results regarding the differences of tidal stream originating from cuspy and cored models. Section 4 describes how these differences can be used to constrain the DM profile of dSphs with associated tidal streams.

\begin{figure}
\centering
 \includegraphics[width=\columnwidth]{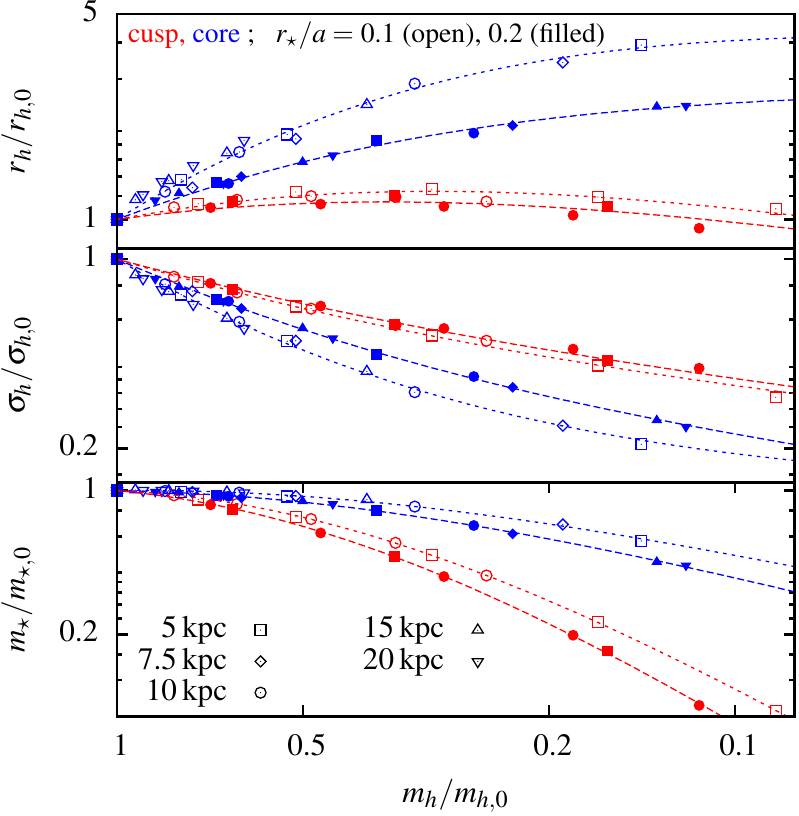}
 \caption{Evolutionary tracks for cuspy and cored dSph galaxies. All quantities are measured at apocentre (see text). The dashed lines are fits of Equation~\ref{equ:trackfit}. The half-light radius $r_h$, the luminosity average velocity dispersion $\sigma_h$ within $r_h$ and the total stellar mass $m_\star$ are normalised to the initial value and are plotted as a function of the current mass $m_h/m_{h,0}$ enclosed within $r_h$. The tracks are independent of the orbit, slightly dependent on the initial stellar segregation $r_\star/a$ and significantly dependent on the shape of the mass profile (cuspy/cored) of the dSph galaxy.}
 \label{fig:tracks}
\end{figure}

\begin{table}
\centering
{
     \caption{Empirical Fit Parameters to the Tidal Evolutionary Tracks}
     \begin{tabular}{llllccc}
     \hline
                             &{${r_\star}/{a} $ }& & &{${r_h}/{r_{h,0}}$ }&{${\sigma_h}/\sigma_{h,0}$ }&{${m_\star}/{m_{\star,0}}$} \\ \hline
     \multirow{4}{*}{\rotatebox[origin=c]{90}{\textsc{Cusp}}}  & \multirow{2}{*}{0.1}      &   $\alpha$ &  & 1.49 & -0.88 & 3.43     \\             
     & & $\beta$  &  & 0.35 & 0.24 & 1.86   \\ 
     & \multirow{2}{*}{0.2} & $\alpha$  & & 1.22 & -0.68 & 3.57   \\
     & & $\beta$ &  & 0.33 & 0.26 & 2.06    \\ \hline
     \multirow{4}{*}{\rotatebox[origin=c]{90}{\textsc{Core}}}  & \multirow{2}{*}{0.1}         &   $\alpha$ &       & 2.91 & -2.56 & 1.43                \\
     & & $\beta$ &  & 0.15 & 0.05 & 0.69   \\
      &\multirow{2}{*}{0.2} & $\alpha$ &  & 1.63 & -1.39 & 0.82   \\
      & & $\beta$  &  & 0.03 & 0.29 & 0.82    \\
      \hline

\end{tabular}
          \label{tab:parameters}
}
\end{table}


\section{Numerical Method}
The \emph{host galaxy} is modelled as a Milky Way-like static potential composed of a spherical NFW halo \citep{nfw1997} with virial radius $r_\mathrm{vir} = \unit[258]{kpc}$, mass $M_\mathrm{vir} = \unit[10^{12}]{M_\odot}$ and concentration $c=12$ \citep{Klypin2002}; 
a spherical Bulge \citep{hernquist1990} with scale radius $a_b = \unit[1.2]{kpc}$ and mass $M_b = \unit[1.3 \cdot 10^{10}]{M_\odot}$,
and an axis-symmetric disk \citep{miyamoto1975} with radial and vertical scale lengths $a_d = \unit[3.5]{kpc}$, $b_d = \unit[0.3]{kpc}$ and mass $M_d = \unit[7.5\cdot10^{10}]{M_\odot}$.
\addtocounter{footnote}{-1}

\begin{figure*}
 \centering
 \includegraphics[width=\textwidth]{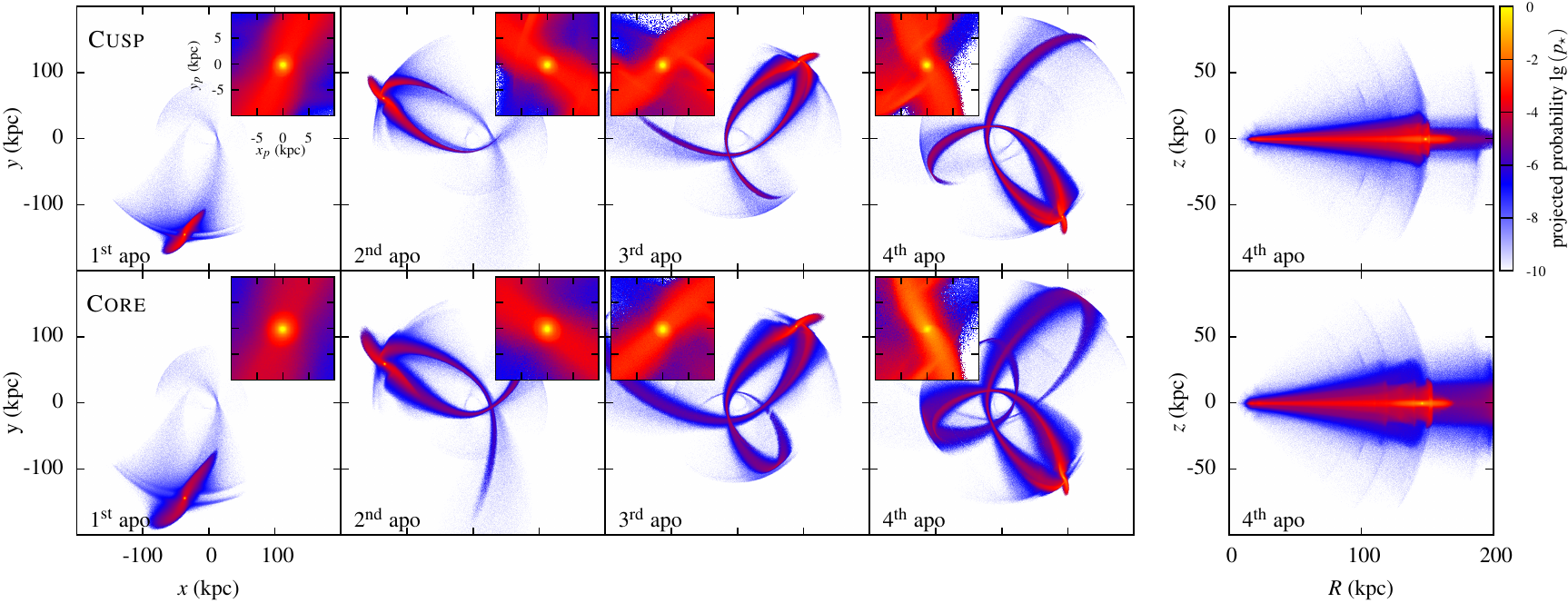}
 \caption{Snapshots of the evolution of tidal streams associated to cuspy (top row) and cored (bottom row) models. Both models have $r_a = \unit[150]{kpc}, r_p = \unit[20]{kpc}, a = \unit[1]{kpc}$ and $r_\star/a = 0.2$. One orbital period equals roughly \unit[2.4]{Gyrs}. The masses have been rescaled to ensure that the luminosity-averaged velocity dispersion within $r_h$ is the same for both progenitors at $t=0$. Small boxes in each panel show a detail view of the progenitor. The rightmost column shows the projection of tidal stream and progenitor on the $\left(R,z\right)$ plane, where $R$ is the cylindrical distance from the galactic centre and $z$ is the distance above the orbital plane of the progenitor. DM particles are colour-coded according to their probability of tagging stars. See \href{http://arxiv.org/src/1501.04968v2/anc}{ArXiv ancillary files} for animations\protect\footnotemark.} 
 \label{fig:streamevol}
\end{figure*}

To generate $N$-body models of dSphs in dynamical equilibrium we draw $2\times 10^7$ particles from a \citet{dehnen1993} distribution function. These models follow an analytical density profile
\begin{equation}
\varrho(r) = \frac{M \left(3-\gamma\right) a}{4\upi r^\gamma \left(r+a\right)^{4-\gamma}} ~,
\end{equation} 
where $\gamma\equiv - \lim_{r \rightarrow 0}{\rm d log} \rho/{\rm d log} r$ defines the slope of the inner density profile. In this work we consider two values, namely $\gamma = 1$ (cusp) and $\gamma = 0$ (core), and set $a = \unit[1]{kpc}$.
Our dSphs models are built under the assumption that stars behave as massless tracers of the underlying DM potential. To generate a stellar component embedded in the DM halo in equilibrium we follow the approach described by \citet{Bullock2005}, where each individual DM particle is assigned a probability, or mass-to-light ratio, of representing a 'star' in such a way that the overall stellar density profile follows a Plummer sphere with segregation $r_\star/a\simeq  r_h/(1.3a)$, where $r_h$ and $r_\star$ are the half-light and core radii of the Plummer model, respectively. To facilitate a comparison between cored and cuspy models the dSph parameters are chosen such that for a given spatial segregation both cored and cuspy DM haloes have the same mass enclosed within the half-light radius $m_h \equiv m(<r_h)$ at $t=0$. Since the mass profile of a Dehnen model is $m(<r)=M[ r/(r+a) ]^{3-\gamma}$, imposing $m_{h,\rm cusp}=m_{h, \rm core}$ requires changing the total mass of the dwarf models by $M_{\rm cusp}/M_{\rm core}=r_h/(r_h+a)$.
Fixing $M_{\rm cusp} = \unit[10^8]{M_\odot}$ yields $M_{\rm core} = 8.66\cdot \unit[10^8]{M_\odot}$ for $r_\star/a =0.1$, and $M_{\rm core} = 4.83\cdot\unit[10^8]{M_\odot}$ for $r_\star/a =0.2$.
Our models are purely gravitational, thus they are scale free, and physical units are just given for clarity.

We use the particle-mesh code \textsc{Superbox} \citep[see][]{Fellhauer2000}, which samples the density in three grids with different resolution and performs a leapfrog integration to solve the equation of motion. In our models each grid consists of $128^3$ cubic cells. Grids 1 and 2 move with the centre of density of the progenitor, while grid 3 is centred on the host galaxy. Grid 1 resolves the core of the progenitor with a resolution of $2a/126 \approx \unit[16]{pc}$, while grids 2 and 3 have resolutions of $20a/126 \approx \unit[160]{pc}$ and $400a/126 \approx \unit[3.2]{kpc}$, respectively. We use a fixed time step of $\unit[1]{Myr}$ and store snapshots every \unit[200]{Myrs}. All models are run in isolation for \unit[14]{Gyrs} in order to test the stability of our numerical set-up. 

The dSph models are placed on highly eccentric orbits with a fixed apocentre distance $r_a = \unit[150]{kpc}$ and pericentres $r_p=5,7.5,10,15$ and \unit[20]{kpc}. These orbits are confined to the plane perpendicular to the symmetry axis of the host disc in order to facilitate the analysis of the associated tidal streams.
As tides strip the dSph models we follow the evolution of the half-light radius $r_h$, the luminosity-averaged velocity dispersion $\sigma_h$ and the stellar mass $m_\star$ as a function of the mass $m_h$ within $r_h$. All quantities reported in this paper are non-projected. Note that $m_h$ can be in principle measured with relatively high accuracy in isolated systems \citep[see e.g.][]{Walker2009, Wolf2010}, but the calculation of $\sigma$ requires a luminosity-averaged integral over the entire dSph, which in case of the dSph undergoing tidal stripping may be difficult to estimate owing to the presence of transient `extra-tidal' features.
To compute the half-light radius $r_h$ and the stellar mass $m_\star$, we fit a Plummer profile to the DM particles weighted by the corresponding mass-to-light ratio. To avoid contamination with `extra-tidal' particles we follow an iterative approach, where only $3/4$ of the particles within a radius of $\unit[5]{kpc}$ to the dSph are fitted at a time. Then using the fitted Plummer parameters we derive the radius that contains $3/4$ of the stellar mass and repeat the fit again. These steps are re-iterated until convergence is reached. The stellar velocity dispersion $\sigma_h$ and mass $m_h$ are computed from the velocities of the particles located within the half-light radius.
 
Because the internal dynamical time of the dSph models are much shorter than the period of the orbit, the progenitor reaches an equilibrium configuration shortly after each pericentre passage. To decrease stochastic fluctuations parameters are averaged over 7 snapshots chosen symmetrically around each apocentre.


\section{Tidal evolutionary tracks}

Previous studies \citep[e.g.][]{penarrubia2008, Penarrubia2010} have shown that tidal effects tend to lower the velocity dispersion and increase the half-light radii of stars embedded within \citet{nfw1997} halos, and that the structural changes in the stellar component of these galaxies follow well-defined {\it evolutionary tracks} that are mainly controlled by how much (and not how) the mass $m_h$ enclosed within the half-light radius $r_h$ varies.
Fig.~\ref{fig:tracks} shows that this result also holds for galaxies embedded in cored DM haloes. However, while the tidal tracks associated to cuspy DM haloes scarcely depend on the spatial segregation of the stellar component within the DM halo, as shown by \citet{penarrubia2008}, those associated to cored DM haloes appear to be sensitive to the initial value of $r_\star/a$. In particular, we find that the half-light radius increases to a larger extent the more deeply segregated the stars are within the DM core.
For the remainder of this paper, it is useful to parametrize the evolutionary tracks using the empirical formula of \citet{penarrubia2008} 
\begin{equation}
f\left(x \right) = \frac{2^\alpha x^\beta}{\left(1+x\right)^\alpha}~~~\mbox{where}~~~x = m_h/m_{h,0}~~.
\label{equ:trackfit}
\end{equation}
The best-fitting parameters are listed in Table \ref{tab:parameters} and the resulting tracks are plotted with dotted and dashed lines in Fig.~\ref{fig:tracks}.
Fig.~\ref{fig:tracks} also shows other points of interest. First, note that at a fixed remnant mass $m_h/m_{h,0}$ cored progenitors have larger half-light radii and colder velocity dispersions than their cuspy counterparts. This difference can be easily understood in terms of the dynamical time of the progenitor galaxies, which scales as $t_{\rm dyn}\propto (G \rho )^{-1/2}$. Because $\rho_{{\rm cusp}}\gg \rho_{{\rm core}}$ at $r\ll a$ we find that stars embedded in DM cusps react adiabatically to external perturbations. In contrast, stars embedded in cored DM haloes tend to absorb a relatively large amount of energy during tidal encounters, which leads to an overall expansion and a low kinematic temperature once dynamical equilibrium is reached.

A second interesting point can be gleaned from the lower panel of Fig.~\ref{fig:tracks}, which shows the fraction of stars lost to tides versus the remnant mass. We find that at a given remnant mass cored dSph models lost {\it less} stars than their cuspy counterparts. This apparently counter-intuitive result arises from the dichotomy between the shapes of the stellar and DM density profiles. Whereas stars in in cored DM halo models tend to be associated to the most-bound DM particles, in cuspy DM haloes a large fraction of particles with high binding energy are located within the density cusp and have a low probability of tagging a star, i.e. a high mass-to-light ratio. E.g. for cuspy models with $r_\star/a = 0.2$, one finds that $90\%$ of the stellar particles are associated to the $14.5\%$ most-bound DM particles, while in cored models this fraction falls to a mere $5\%$. Note, however, that because dSphs embedded in cored DM haloes are in general less resilient to tidal stripping, these models tend to evolve at a faster pace along the evolutionary tracks than those embedded in cuspy haloes \citep[e.g.][]{Penarrubia2010}.

\begin{figure}
\centering
\includegraphics[width=0.89\columnwidth]{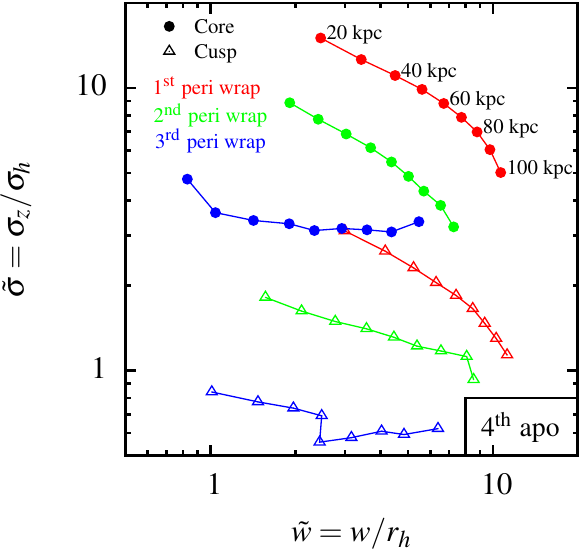}
 \caption{Width $\tilde w$ and velocity dispersion $\tilde \sigma$ of the streams normalized to the half-light radius and luminosity-averaged velocity dispersion of the progenitor dSph, respectively. We show data at a fixed snapshot (4\textsuperscript{th} apocentre) of the simulation shown in Fig.~\ref{fig:streamevol}. We distinguish between dSph models embedded in cuspy (open symbols) and cored (filled symbols) DM haloes. Symbols are colour-coded according to the pericentric passage at which particles became tidally unbound. We mark the cylindrical distance $R$ to the galactic centre of segments within the range $\unit[20]{kpc} \leq R \leq \unit[100]{kpc}$. Notice the clear kinematic offset of the streams depending on the DM halo profile.}
 \label{fig:sigmawidth}
\end{figure}


\section{Stream evolution}
\footnotetext{\href{http://arxiv.org/src/1501.04968v2/anc/dSph_stream_brightness.avi}{\textbf{dSph\_stream\_brightness.avi}}: colour-coded by surface brightness \\ \hspace*{1.925mm} \href{http://arxiv.org/src/1501.04968v2/anc/dSph_stream_dispersion.avi}{\textbf{dSph\_stream\_dispersion.avi}}: colour-coded by velocity dispersion}
In this section we compare the properties of stellar streams associated to cored and cuspy progenitors. As expected, Fig.~\ref{fig:streamevol} shows that tides are strongest at pericenter and trigger recurring episodes of mass loss at every pericentric passage. Comparison between the upper and lower panel reveals that tidal mass loss is more pronounced in dSphs embedded in cored DM haloes. Indeed, although not shown here the cored dSph is fully disrupted during the 5th encounter, whereas the cuspy dSph still retains a bound remnant.

At each snapshot we quantify the mass $m_h/m_{h,0}$. The width $w^2 = \langle z^2 \rangle - \langle z \rangle^2$ and the velocity dispersion $\sigma_z$ of particles moving along the tidal tails are measured in the direction perpendicular to the orbital plane at fixed cylindrical distances $R=\sqrt{x^2 +y^2}$ from the host galaxy centre. These quantities are then normalized by the half-light radius and the velocity dispersion of the stream progenitor as given by Equation~2 for the corresponding value of $m_h/m_{h,0}$.
To reduce noise we apply the same averaging is it was done with the tracks. Fig.~\ref{fig:sigmawidth} shows that tidal streams become very narrow and kinematically hot at small distances from the host and get wider and colder with increasing $R$, as also shown in Fig.~\ref{fig:streamevol}, as one would expect from Liouville's theorem. 

In order to understand the internal evolution of the stream properties it is useful to distinguish between particles lost at different pericentric passages. We define a \emph{wrap} as the ensemble of particles stripped at the same pericentre encounter. Each wrap is then divided into \emph{segments} according to their cylindrical distance $R$ to the galactic centre. Wraps that formed during recent pericentric passages tend to be found at close distances from the progenitor galaxy, as it takes a few orbital periods for these particles to populate the entire orbit. In contrast stars stripped at earlier times can in some cases wrap around the host galaxy entirely.
Comparison between different wraps in Fig.~\ref{fig:sigmawidth} shows that streams associated to dSphs embedded in cored DM haloes have similar widths but higher velocity dispersions than their cuspy counterparts. This is so regardless of the galactocentric radii at which these quantities are measured. Also, wraps that form during early encounters tend to have higher velocity dispersions than those formed recently. This trend arises because the velocity dispersion of the dwarf galaxy decreases at a faster pace (see Fig.~\ref{fig:tracks}) than the velocity dispersion of the associated stream. Hence, tidal streams appear kinematically {\it hotter} with respect to to the progenitor's velocity dispersion as they become {\it older}\footnote{Stream `age' is typically defined as the time elapsed since the particles become tidally unbound.}.
\begin{figure}
\centering
\includegraphics[width=0.89\columnwidth]{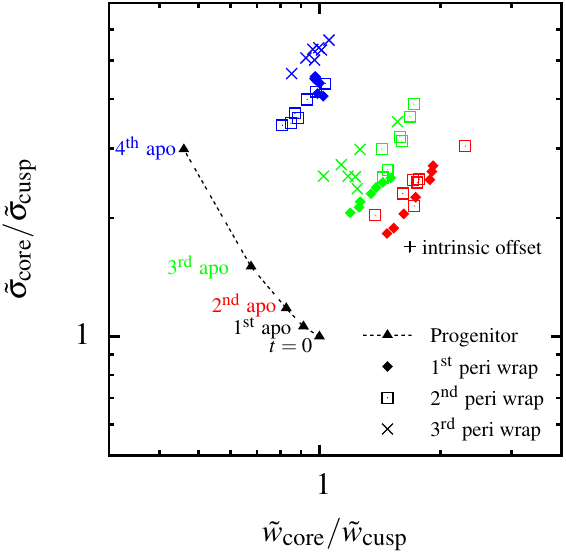}
 \caption{Time evolution of the width and velocity dispersion ratios between streams associated to dSphs embedded in cored and cuspy dark matter haloes of the simulation shown in Fig.~\ref{fig:streamevol}. Symbols denote particles lost at different pericentric passages (wraps), whereas colours distinguish between different snapshots. The width and velocity dispersion ratios roughly follow the ratio between the tidal evolutionary tracks of cuspy and cored dSphs as given by Equation~2, which we plot with a dotted line for ease of reference. }
 \label{fig:sigmawidth2nd}
\end{figure}
Fig.~\ref{fig:sigmawidth} shows that both the width and the velocity dispersion of tidal streams are highly dependent on the progenitor's orbit, the number of orbital periods completed, and the galactocentric radius where these quantities are measured.

Interestingly, Fig.~\ref{fig:sigmawidth2nd} shows that plotting the ratios $\tilde \sigma_{\rm core}/\tilde \sigma_{\rm cusp}$ and $\tilde w_{\rm core}/\tilde w_{\rm cusp}$ at fixed segments collapses the degeneracy between these effects into a single relation, which roughly follows the path defined by the tidal evolutionary tracks of the progenitor galaxies, Equation~2, evaluated for the corresponding values of $m_h/m_{h,0}$ for the cuspy and cored models (shown with dotted lines for ease of reference). This result suggests that the evolutionary path of the stream in this plane is mainly driven by the disparate structural evolution of galaxies embedded in cored and cuspy DM haloes as a result of tidal mass loss (see Fig.~ \ref{fig:tracks}). As shown in Fig.~\ref{fig:sigmawidth2nd}, there is an intrinsic offset between the various properties of the stream which can be understood using simple dynamical arguments. If we assume, as in \citet{Amorisco2014}, that the stream width is proportional to the tidal radius of the progenitor at pericentre $r_t\propto M^{1/3}$, and that the velocity dispersion of the stream scales as $\sigma_z \propto \sqrt{m(<r_t) / r_t} $, one finds that for the orbits plotted in Fig.~\ref{fig:sigmawidth2nd} the width and velocity dispersion ratios are $r_{t, {\rm core}}/r_{t,{\rm cusp}}\simeq 1.7 $ and $\sigma_{z,{\rm core}}/\sigma_{z,{\rm cusp}}\simeq 1.7 $, respectively (see section 2). Fig.~\ref{fig:sigmawidth2nd} shows that these simple analytical estimates, marked with a black cross for ease of reference, faithfully describe the numerical results.
It is straightforward to show that for models with smaller stellar segregation $r_\star/a$ with the same values of $r_h$ and $\sigma_h$ this offset increases. This is in agreement with our models for $r_\star/a=0.1$ (not shown here).

\section{Conclusions}
We have used $N$-body simulations to follow the evolution of dSphs as driven by galactic tides. Our models adopt cosmologically-motivated mass profiles where dSphs are embedded in extended dark matter haloes with density profiles that have either a centrally-divergent `cusp' or an homogeneous-density `core'. As in previous studies, we find that the half-light radius $r_h$, velocity dispersion $\sigma_h$ and stellar mass $m_\star$ of 
dSphs acted on by tides follow well-defined evolutionary tracks that are mainly controlled by the remnant mass $m_h/m_{h,0}$ enclosed within the half-light radius. Crucially, these tracks also appear to be remarkably sensitive to the adopted shape of the halo mass profile (cusp/core), and also to a minor extent to the degree of segregation of stars within the dark matter halo.

Our models indicate that the differences in the evolutionary paths are reflected upon the internal dynamics of the associated tidal streams. In particular, after normalizing the stream width and velocity dispersion by the half-light radius and internal dispersion of the remnant progenitor, respectively, we find that {\it tidal streams of dSphs embedded in cored dark matter models tend to be systematically hotter than their cuspy counterparts}. The kinematic offset 
arises during pericentric encounters and becomes more pronounced as the age of the stream increases.

The results presented in this Letter call for a simple method for inferring the distribution of dark matter in dSphs with associated stellar streams. First, one would derive $m_\star/m_{\star,0}$ by measuring the stellar mass in the stellar stream and in the remnant galaxy. For a given dark matter halo profile and a given initial stellar segregation this observation will fix the position of the galaxy on the tidal tracks shown in Fig~\ref{fig:tracks}, and thus the properties of the progenitor system prior to tidal stripping. Models that reproduce the observed $r_h$ and $\sigma_h$ lead to $M_\mathrm{core} > M_\mathrm{cusp}$.  Having set the initial conditions of the dwarf galaxy, the next step corresponds to modelling the orbit and the mass evolution of the progenitor system simultaneously in a given host potential. This can be done with the aid of self-consistent N-body models \citep[e.g.][]{Bonaca2014} or, alternatively, with efficient semi-analytical algorithms based on forward modelling of stellar streams \citep[e.g.][]{Varghese2011,Sanders2013,Amorisco2014,Gibbons2014,Fardal2014,Price-Whelan2014}. Our results suggest that the velocity dispersion of stellar streams associated to cuspy halo models will lie systematically below that of the cored models, which in turn can be used to constrain the mass profile of the progenitor galaxy by comparing the velocity dispersion of the stream against that of the remnant galaxy.

To date the only dwarf spheroidal with a known tidal stream corresponds to the Sagittarius dSph \citetext{\citealp{Ibata1997}, see \citealp{Belokurov2014} for a recent data compilation.} As this dSph is currently located close to pericentre, we can't apply directly our tidal evolutionary tracks, as these have been computed for dSphs that have reached equilibrium after each pericentre passage. The use of $N$-body simulations will be required to study perturbations from the equilibrium configuration. An application of the proposed method to the existing kinematic data set will be presented in a separate contribution.

\footnotesize{
\bibliography{TidalStreams}

\begin{thebibliography}{}
\makeatletter
\relax
\def\mn@urlcharsother{\let\do\@makeother \do\$\do\&\do\#\do\^\do\_\do\%\do\~}
\def\mn@doi{\begingroup\mn@urlcharsother \@ifnextchar [ {\mn@doi@}
  {\mn@doi@[]}}
\def\mn@doi@[#1]#2{\def\@tempa{#1}\ifx\@tempa\@empty \href
  {http://dx.doi.org/#2} {doi:#2}\else \href {http://dx.doi.org/#2} {#1}\fi
  \endgroup}
\def\mn@eprint#1#2{\mn@eprint@#1:#2::\@nil}
\def\mn@eprint@arXiv#1{\href {http://arxiv.org/abs/#1} {{arXiv:#1}}}
\def\mn@eprint@dblp#1{\href {http://dblp.uni-trier.de/rec/bibtex/#1.xml}
  {dblp:#1}}
\def\mn@eprint@#1:#2:#3:#4\@nil{\def\@tempa {#1}\def\@tempb {#2}\def\@tempc
  {#3}\ifx \@tempc \@empty \let \@tempc \@tempb \let \@tempb \@tempa \fi \ifx
  \@tempb \@empty \def\@tempb {arXiv}\fi \@ifundefined
  {mn@eprint@\@tempb}{\@tempb:\@tempc}{\expandafter \expandafter \csname
  mn@eprint@\@tempb\endcsname \expandafter{\@tempc}}}

\bibitem[\protect\citeauthoryear{{Amorisco}}{{Amorisco}}{2014}]{Amorisco2014}
{Amorisco} N.~C.,  2014, preprint, \href
  {http://adsabs.harvard.edu/abs/2014arXiv1410.0360A} {} (\mn@eprint {arXiv}
  {1410.0360})

\bibitem[\protect\citeauthoryear{{Amorisco} \& {Evans}}{{Amorisco} \&
  {Evans}}{2012}]{Amorisco2012}
{Amorisco} N.~C.,  {Evans} N.~W.,  2012, \mn@doi [\mnras]
  {10.1111/j.1365-2966.2011.19684.x}, \href
  {http://adsabs.harvard.edu/abs/2012MNRAS.419..184A} {419, 184}

\bibitem[\protect\citeauthoryear{{Battaglia}, {Helmi}, {Tolstoy}, {Irwin},
  {Hill}  \& {Jablonka}}{{Battaglia} et~al.}{2008}]{Battaglia2008}
{Battaglia} G.,  {Helmi} A.,  {Tolstoy} E.,  {Irwin} M.,  {Hill} V.,
  {Jablonka} P.,  2008, \mn@doi [\apjl] {10.1086/590179}, \href
  {http://adsabs.harvard.edu/abs/2008ApJ...681L..13B} {681, L13}

\bibitem[\protect\citeauthoryear{{Belokurov} et~al.,}{{Belokurov}
  et~al.}{2014}]{Belokurov2014}
{Belokurov} V.,  et~al., 2014, \mn@doi [\mnras] {10.1093/mnras/stt1862}, \href
  {http://adsabs.harvard.edu/abs/2014MNRAS.437..116B} {437, 116}

\bibitem[\protect\citeauthoryear{{Bonaca}, {Geha}, {K{\"u}pper}, {Diemand},
  {Johnston}  \& {Hogg}}{{Bonaca} et~al.}{2014}]{Bonaca2014}
{Bonaca} A.,  {Geha} M.,  {K{\"u}pper} A.~H.~W.,  {Diemand} J.,  {Johnston}
  K.~V.,   {Hogg} D.~W.,  2014, \mn@doi [\apj] {10.1088/0004-637X/795/1/94},
  \href {http://adsabs.harvard.edu/abs/2014ApJ...795...94B} {795, 94}

\bibitem[\protect\citeauthoryear{{Boyarsky}, {Ruchayskiy}  \&
  {Iakubovskyi}}{{Boyarsky} et~al.}{2009}]{Boyarsky2009}
{Boyarsky} A.,  {Ruchayskiy} O.,   {Iakubovskyi} D.,  2009, \mn@doi [\jcap]
  {10.1088/1475-7516/2009/03/005}, \href
  {http://adsabs.harvard.edu/abs/2009JCAP...03..005B} {3, 5}

\bibitem[\protect\citeauthoryear{{Breddels} \& {Helmi}}{{Breddels} \&
  {Helmi}}{2013}]{Breddels2013}
{Breddels} M.~A.,  {Helmi} A.,  2013, \mn@doi [\aap]
  {10.1051/0004-6361/201321606}, \href
  {http://adsabs.harvard.edu/abs/2013A%26A...558A..35B} {558, A35}

\bibitem[\protect\citeauthoryear{{Brooks} \& {Zolotov}}{{Brooks} \&
  {Zolotov}}{2014}]{Brooks2014}
{Brooks} A.~M.,  {Zolotov} A.,  2014, \mn@doi [\apj]
  {10.1088/0004-637X/786/2/87}, \href
  {http://adsabs.harvard.edu/abs/2014ApJ...786...87B} {786, 87}

\bibitem[\protect\citeauthoryear{{Bullock} \& {Johnston}}{{Bullock} \&
  {Johnston}}{2005}]{Bullock2005}
{Bullock} J.~S.,  {Johnston} K.~V.,  2005, \mn@doi [ApJ] {10.1086/497422},
  \href {http://adsabs.harvard.edu/abs/2005ApJ...635..931B} {635, 931}

\bibitem[\protect\citeauthoryear{{Burkert}}{{Burkert}}{2000}]{Burkert2000}
{Burkert} A.,  2000, \mn@doi [\apjl] {10.1086/312674}, \href
  {http://adsabs.harvard.edu/abs/2000ApJ...534L.143B} {534, L143}

\bibitem[\protect\citeauthoryear{{Dehnen}}{{Dehnen}}{1993}]{dehnen1993}
{Dehnen} W.,  1993, MNRAS, \href
  {http://adsabs.harvard.edu/abs/1993MNRAS.265..250D} {265, 250}

\bibitem[\protect\citeauthoryear{{Di Cintio}, {Brook}, {Macci{\`o}}, {Stinson},
  {Knebe}, {Dutton}  \& {Wadsley}}{{Di Cintio} et~al.}{2014}]{DiCintio2014}
{Di Cintio} A.,  {Brook} C.~B.,  {Macci{\`o}} A.~V.,  {Stinson} G.~S.,  {Knebe}
  A.,  {Dutton} A.~A.,   {Wadsley} J.,  2014, \mn@doi [\mnras]
  {10.1093/mnras/stt1891}, \href
  {http://adsabs.harvard.edu/abs/2014MNRAS.437..415D} {437, 415}

\bibitem[\protect\citeauthoryear{{Eyre} \& {Binney}}{{Eyre} \&
  {Binney}}{2011}]{Eyre2011}
{Eyre} A.,  {Binney} J.,  2011, \mn@doi [\mnras]
  {10.1111/j.1365-2966.2011.18270.x}, \href
  {http://adsabs.harvard.edu/abs/2011MNRAS.413.1852E} {413, 1852}

\bibitem[\protect\citeauthoryear{{Fardal}, {Huang}  \& {Weinberg}}{{Fardal}
  et~al.}{2014}]{Fardal2014}
{Fardal} M.~A.,  {Huang} S.,   {Weinberg} M.~D.,  2014, preprint, \href
  {http://adsabs.harvard.edu/abs/2014arXiv1410.1861F} {} (\mn@eprint {arXiv}
  {1410.1861})

\bibitem[\protect\citeauthoryear{{Fellhauer}, {Kroupa}, {Baumgardt}, {Bien},
  {Boily}, {Spurzem}  \& {Wassmer}}{{Fellhauer} et~al.}{2000}]{Fellhauer2000}
{Fellhauer} M.,  {Kroupa} P.,  {Baumgardt} H.,  {Bien} R.,  {Boily} C.~M.,
  {Spurzem} R.,   {Wassmer} N.,  2000, \mn@doi [NA]
  {10.1016/S1384-1076(00)00032-4}, \href
  {http://adsabs.harvard.edu/abs/2000NewA....5..305F} {5, 305}

\bibitem[\protect\citeauthoryear{{Gibbons}, {Belokurov}  \& {Evans}}{{Gibbons}
  et~al.}{2014}]{Gibbons2014}
{Gibbons} S.~L.~J.,  {Belokurov} V.,   {Evans} N.~W.,  2014, \mn@doi [\mnras]
  {10.1093/mnras/stu1986}, \href
  {http://adsabs.harvard.edu/abs/2014MNRAS.445.3788G} {445, 3788}

\bibitem[\protect\citeauthoryear{{Gilmore}, {Wilkinson}, {Wyse}, {Kleyna},
  {Koch}, {Evans}  \& {Grebel}}{{Gilmore} et~al.}{2007}]{Gilmore2007}
{Gilmore} G.,  {Wilkinson} M.~I.,  {Wyse} R.~F.~G.,  {Kleyna} J.~T.,  {Koch}
  A.,  {Evans} N.~W.,   {Grebel} E.~K.,  2007, \mn@doi [ApJ] {10.1086/518025},
  \href {http://adsabs.harvard.edu/abs/2007ApJ...663..948G} {663, 948}

\bibitem[\protect\citeauthoryear{{Hernquist}}{{Hernquist}}{1990}]{hernquist1990}
{Hernquist} L.,  1990, \mn@doi [ApJ] {10.1086/168845}, \href
  {http://adsabs.harvard.edu/abs/1990ApJ...356..359H} {356, 359}

\bibitem[\protect\citeauthoryear{{Ibata}, {Wyse}, {Gilmore}, {Irwin}  \&
  {Suntzeff}}{{Ibata} et~al.}{1997}]{Ibata1997}
{Ibata} R.~A.,  {Wyse} R.~F.~G.,  {Gilmore} G.,  {Irwin} M.~J.,   {Suntzeff}
  N.~B.,  1997, \mn@doi [\aj] {10.1086/118283}, \href
  {http://adsabs.harvard.edu/abs/1997AJ....113..634I} {113, 634}

\bibitem[\protect\citeauthoryear{{Klypin}, {Zhao}  \& {Somerville}}{{Klypin}
  et~al.}{2002}]{Klypin2002}
{Klypin} A.,  {Zhao} H.,   {Somerville} R.~S.,  2002, \mn@doi [ApJ]
  {10.1086/340656}, \href {http://adsabs.harvard.edu/abs/2002ApJ...573..597K}
  {573, 597}

\bibitem[\protect\citeauthoryear{{K{\"u}pper}, {MacLeod}  \&
  {Heggie}}{{K{\"u}pper} et~al.}{2008}]{Kupper2008}
{K{\"u}pper} A.~H.~W.,  {MacLeod} A.,   {Heggie} D.~C.,  2008, \mn@doi [\mnras]
  {10.1111/j.1365-2966.2008.13323.x}, \href
  {http://adsabs.harvard.edu/abs/2008MNRAS.387.1248K} {387, 1248}

\bibitem[\protect\citeauthoryear{{K{\"u}pper}, {Lane}  \&
  {Heggie}}{{K{\"u}pper} et~al.}{2012}]{Kupper2012}
{K{\"u}pper} A.~H.~W.,  {Lane} R.~R.,   {Heggie} D.~C.,  2012, \mn@doi [\mnras]
  {10.1111/j.1365-2966.2011.20242.x}, \href
  {http://adsabs.harvard.edu/abs/2012MNRAS.420.2700K} {420, 2700}

\bibitem[\protect\citeauthoryear{{Loeb} \& {Weiner}}{{Loeb} \&
  {Weiner}}{2011}]{Loeb2011}
{Loeb} A.,  {Weiner} N.,  2011, \mn@doi [Physical Review Letters]
  {10.1103/PhysRevLett.106.171302}, \href
  {http://adsabs.harvard.edu/abs/2011PhRvL.106q1302L} {106, 171302}

\bibitem[\protect\citeauthoryear{{Lovell}, {Frenk}, {Eke}, {Jenkins}, {Gao}  \&
  {Theuns}}{{Lovell} et~al.}{2014}]{Lovell2014}
{Lovell} M.~R.,  {Frenk} C.~S.,  {Eke} V.~R.,  {Jenkins} A.,  {Gao} L.,
  {Theuns} T.,  2014, \mn@doi [\mnras] {10.1093/mnras/stt2431}, \href
  {http://adsabs.harvard.edu/abs/2014MNRAS.439..300L} {439, 300}

\bibitem[\protect\citeauthoryear{{Macci{\`o}}, {Paduroiu}, {Anderhalden},
  {Schneider}  \& {Moore}}{{Macci{\`o}} et~al.}{2012}]{Maccio2012}
{Macci{\`o}} A.~V.,  {Paduroiu} S.,  {Anderhalden} D.,  {Schneider} A.,
  {Moore} B.,  2012, \mn@doi [\mnras] {10.1111/j.1365-2966.2012.21284.x}, \href
  {http://adsabs.harvard.edu/abs/2012MNRAS.424.1105M} {424, 1105}

\bibitem[\protect\citeauthoryear{{McConnachie}}{{McConnachie}}{2012}]{McConnachie2012}
{McConnachie} A.~W.,  2012, \mn@doi [\aj] {10.1088/0004-6256/144/1/4}, \href
  {http://adsabs.harvard.edu/abs/2012AJ....144....4M} {144, 4}

\bibitem[\protect\citeauthoryear{{Miyamoto} \& {Nagai}}{{Miyamoto} \&
  {Nagai}}{1975}]{miyamoto1975}
{Miyamoto} M.,  {Nagai} R.,  1975, PASJ, \href
  {http://adsabs.harvard.edu/abs/1975PASJ...27..533M} {27, 533}

\bibitem[\protect\citeauthoryear{{Navarro}, {Eke}  \& {Frenk}}{{Navarro}
  et~al.}{1996}]{Navarro96}
{Navarro} J.~F.,  {Eke} V.~R.,   {Frenk} C.~S.,  1996, \mnras, \href
  {http://adsabs.harvard.edu/abs/1996MNRAS.283L..72N} {283, L72}

\bibitem[\protect\citeauthoryear{{Navarro}, {Frenk}  \& {White}}{{Navarro}
  et~al.}{1997}]{nfw1997}
{Navarro} J.~F.,  {Frenk} C.~S.,   {White} S.~D.~M.,  1997, \mn@doi [ApJ]
  {10.1086/304888}, \href {http://adsabs.harvard.edu/abs/1997ApJ...490..493N}
  {490, 493}

\bibitem[\protect\citeauthoryear{{Nipoti} \& {Binney}}{{Nipoti} \&
  {Binney}}{2015}]{Nipoti2014}
{Nipoti} C.,  {Binney} J.,  2015, \mn@doi [\mnras] {10.1093/mnras/stu2217},
  \href {http://adsabs.harvard.edu/abs/2015MNRAS.446.1820N} {446, 1820}

\bibitem[\protect\citeauthoryear{{Pe{\~n}arrubia}, {Benson},
  {Mart{\'{\i}}nez-Delgado}  \& {Rix}}{{Pe{\~n}arrubia}
  et~al.}{2006}]{Penarrubia2006}
{Pe{\~n}arrubia} J.,  {Benson} A.~J.,  {Mart{\'{\i}}nez-Delgado} D.,   {Rix}
  H.~W.,  2006, \mn@doi [\apj] {10.1086/504316}, \href
  {http://adsabs.harvard.edu/abs/2006ApJ...645..240P} {645, 240}

\bibitem[\protect\citeauthoryear{{Pe{\~n}arrubia}, {Navarro}  \&
  {McConnachie}}{{Pe{\~n}arrubia} et~al.}{2008}]{penarrubia2008}
{Pe{\~n}arrubia} J.,  {Navarro} J.~F.,   {McConnachie} A.~W.,  2008, \mn@doi
  [ApJ] {10.1086/523686}, \href
  {http://adsabs.harvard.edu/abs/2008ApJ...673..226P} {673, 226}

\bibitem[\protect\citeauthoryear{{Pe{\~n}arrubia}, {Benson}, {Walker},
  {Gilmore}, {McConnachie}  \& {Mayer}}{{Pe{\~n}arrubia}
  et~al.}{2010}]{Penarrubia2010}
{Pe{\~n}arrubia} J.,  {Benson} A.~J.,  {Walker} M.~G.,  {Gilmore} G.,
  {McConnachie} A.~W.,   {Mayer} L.,  2010, \mn@doi [MNRAS]
  {10.1111/j.1365-2966.2010.16762.x}, \href
  {http://adsabs.harvard.edu/abs/2010MNRAS.406.1290P} {406, 1290}

\bibitem[\protect\citeauthoryear{{Pe{\~n}arrubia}, {Pontzen}, {Walker}  \&
  {Koposov}}{{Pe{\~n}arrubia} et~al.}{2012}]{Penarrubia2012}
{Pe{\~n}arrubia} J.,  {Pontzen} A.,  {Walker} M.~G.,   {Koposov} S.~E.,  2012,
  \mn@doi [\apjl] {10.1088/2041-8205/759/2/L42}, \href
  {http://adsabs.harvard.edu/abs/2012ApJ...759L..42P} {759, L42}

\bibitem[\protect\citeauthoryear{{Pontzen} \& {Governato}}{{Pontzen} \&
  {Governato}}{2012}]{Pontzen2012}
{Pontzen} A.,  {Governato} F.,  2012, \mn@doi [\mnras]
  {10.1111/j.1365-2966.2012.20571.x}, \href
  {http://adsabs.harvard.edu/abs/2012MNRAS.421.3464P} {421, 3464}

\bibitem[\protect\citeauthoryear{{Pontzen} \& {Governato}}{{Pontzen} \&
  {Governato}}{2014}]{Pontzen2014}
{Pontzen} A.,  {Governato} F.,  2014, \mn@doi [\nat] {10.1038/nature12953},
  \href {http://adsabs.harvard.edu/abs/2014Natur.506..171P} {506, 171}

\bibitem[\protect\citeauthoryear{{Price-Whelan}, {Hogg}, {Johnston}  \&
  {Hendel}}{{Price-Whelan} et~al.}{2014}]{Price-Whelan2014}
{Price-Whelan} A.~M.,  {Hogg} D.~W.,  {Johnston} K.~V.,   {Hendel} D.,  2014,
  \mn@doi [\apj] {10.1088/0004-637X/794/1/4}, \href
  {http://adsabs.harvard.edu/abs/2014ApJ...794....4P} {794, 4}

\bibitem[\protect\citeauthoryear{{Richardson} \& {Fairbairn}}{{Richardson} \&
  {Fairbairn}}{2014}]{Richardson2014}
{Richardson} T.,  {Fairbairn} M.,  2014, \mn@doi [\mnras]
  {10.1093/mnras/stu691}, \href
  {http://adsabs.harvard.edu/abs/2014MNRAS.441.1584R} {441, 1584}

\bibitem[\protect\citeauthoryear{{Rocha}, {Peter}, {Bullock}, {Kaplinghat},
  {Garrison-Kimmel}, {O{\~n}orbe}  \& {Moustakas}}{{Rocha}
  et~al.}{2013}]{Rocha2013}
{Rocha} M.,  {Peter} A.~H.~G.,  {Bullock} J.~S.,  {Kaplinghat} M.,
  {Garrison-Kimmel} S.,  {O{\~n}orbe} J.,   {Moustakas} L.~A.,  2013, \mn@doi
  [\mnras] {10.1093/mnras/sts514}, \href
  {http://adsabs.harvard.edu/abs/2013MNRAS.430...81R} {430, 81}

\bibitem[\protect\citeauthoryear{{Sanders} \& {Binney}}{{Sanders} \&
  {Binney}}{2013}]{Sanders2013}
{Sanders} J.~L.,  {Binney} J.,  2013, \mn@doi [\mnras] {10.1093/mnras/stt816},
  \href {http://adsabs.harvard.edu/abs/2013MNRAS.433.1826S} {433, 1826}

\bibitem[\protect\citeauthoryear{{Strigari}, {Frenk}  \& {White}}{{Strigari}
  et~al.}{2014}]{Strigari2014}
{Strigari} L.~E.,  {Frenk} C.~S.,   {White} S.~D.~M.,  2014, preprint, \href
  {http://adsabs.harvard.edu/abs/2014arXiv1406.6079S} {} (\mn@eprint {arXiv}
  {1406.6079})

\bibitem[\protect\citeauthoryear{{Tremaine} \& {Gunn}}{{Tremaine} \&
  {Gunn}}{1979}]{TremaineGunn1979}
{Tremaine} S.,  {Gunn} J.~E.,  1979, \mn@doi [Physical Review Letters]
  {10.1103/PhysRevLett.42.407}, \href
  {http://adsabs.harvard.edu/abs/1979PhRvL..42..407T} {42, 407}

\bibitem[\protect\citeauthoryear{{Varghese}, {Ibata}  \& {Lewis}}{{Varghese}
  et~al.}{2011}]{Varghese2011}
{Varghese} A.,  {Ibata} R.,   {Lewis} G.~F.,  2011, \mn@doi [\mnras]
  {10.1111/j.1365-2966.2011.19097.x}, \href
  {http://adsabs.harvard.edu/abs/2011MNRAS.417..198V} {417, 198}

\bibitem[\protect\citeauthoryear{{Vogelsberger}, {Zavala}  \&
  {Loeb}}{{Vogelsberger} et~al.}{2012}]{Vogelsberger2012}
{Vogelsberger} M.,  {Zavala} J.,   {Loeb} A.,  2012, \mn@doi [\mnras]
  {10.1111/j.1365-2966.2012.21182.x}, \href
  {http://adsabs.harvard.edu/abs/2012MNRAS.423.3740V} {423, 3740}

\bibitem[\protect\citeauthoryear{{Vogelsberger}, {Zavala}, {Simpson}  \&
  {Jenkins}}{{Vogelsberger} et~al.}{2014}]{Vogelsberger2014}
{Vogelsberger} M.,  {Zavala} J.,  {Simpson} C.,   {Jenkins} A.,  2014, \mn@doi
  [\mnras] {10.1093/mnras/stu1713}, \href
  {http://adsabs.harvard.edu/abs/2014MNRAS.444.3684V} {444, 3684}

\bibitem[\protect\citeauthoryear{{Walker} \& {Pe{\~n}arrubia}}{{Walker} \&
  {Pe{\~n}arrubia}}{2011}]{Walker2011}
{Walker} M.~G.,  {Pe{\~n}arrubia} J.,  2011, \mn@doi [ApJ]
  {10.1088/0004-637X/742/1/20}, \href
  {http://adsabs.harvard.edu/abs/2011ApJ...742...20W} {742, 20}

\bibitem[\protect\citeauthoryear{{Walker}, {Mateo}, {Olszewski},
  {Pe{\~n}arrubia}, {Wyn Evans}  \& {Gilmore}}{{Walker}
  et~al.}{2009}]{Walker2009}
{Walker} M.~G.,  {Mateo} M.,  {Olszewski} E.~W.,  {Pe{\~n}arrubia} J.,  {Wyn
  Evans} N.,   {Gilmore} G.,  2009, \mn@doi [ApJ]
  {10.1088/0004-637X/704/2/1274}, \href
  {http://adsabs.harvard.edu/abs/2009ApJ...704.1274W} {704, 1274}

\bibitem[\protect\citeauthoryear{{Wolf}, {Martinez}, {Bullock}, {Kaplinghat},
  {Geha}, {Mu{\~n}oz}, {Simon}  \& {Avedo}}{{Wolf} et~al.}{2010}]{Wolf2010}
{Wolf} J.,  {Martinez} G.~D.,  {Bullock} J.~S.,  {Kaplinghat} M.,  {Geha} M.,
  {Mu{\~n}oz} R.~R.,  {Simon} J.~D.,   {Avedo} F.~F.,  2010, \mn@doi [MNRAS]
  {10.1111/j.1365-2966.2010.16753.x}, \href
  {http://adsabs.harvard.edu/abs/2010MNRAS.406.1220W} {406, 1220}

\makeatother
\end{thebibliography}
}

\label{lastpage}

\end{document}